\documentstyle[aps,preprint]{revtex}
\tightenlines

\begin{document}

\def\dx{\displaystyle}
\def\ang#1{\left \langle #1 \right \rangle}
\def\angg#1{\left \langle \left \langle  #1 \right \rangle  \right
\rangle} \def\const{\hbox{\rm const}}
\def\be{\begin{equation}}
\def\ee{\end{equation}}
\def\f{\frac}
\def\p{\partial}
\def\eps{\varepsilon}

\title { Comment on "Quantum diffusion of $^3$ He impurities in solid $^4$ He"
 }
\author{                  Dimitar I. Pushkarov}
\address{                 Institute of Solid State Physics
                        Bulgarian Academy of Sciences,
                            Sofia 1784, Bulgaria,\\
                          E-mail: dipushk@issp.bas.bg
 }
\maketitle
\begin{abstract}
      In this comment\footnote{This Comment was sent to the Editor of 
Phys.Rev.B on 03.12.1998, accepted for publication in August 1999, but not published till now.} I show that the experimental data on quantum diffusion
of $^3$He impurities in solid $^4$He can be explained using the adopted
quasiparticle theory. The contention by E.G. Kisvarsanyi and N.S. Sullivan
(KS) in  Phys.Rev.B {\bf 48} 16557 (1993) as well as in their Reply (ibid. {\bf
55} 3989 (1997))  to the Grigor'ev's Comment (Phys.Rev.B {\bf 55} 3987 (1997))
that "Pushkarov's theory of phonon scattering fails to fit the data by very
large factors" is groundless and may result from their bad arithmetical error.
This means that the phonon-impurity scattering mechanism of diffusion is
consistent with experiment and its neglecting by KS makes their results
questionable.
\end{abstract}

        The temperature dependence of the impuriton
diffusion coefficient in solid helium can be determined by two types of thermal
excitations -- phonons and vacancies. Both of them have been considered in a
number of works (see e.g. \cite{Allen,Locke}). For low concentrations and
temperatures $T\le 1$ K the theory based on
the impuriton--phonon scattering has been confirmed by the experiment (cf e.g.
\cite{Allen,Mikheev,Moscow} and the references therein) and used for
determination of quasiparticle characteristics. The quantitative
agreement was first obtained in Ref. \cite{Pisma} (cf also \cite{FNT,P1975}),
and since that time the theory has not been subjected to principle changes.

        Kisvarsanyi and Sullivan (KS) argue in their work \cite{KS93} and in
their Reply \cite{Reply} to the Grigor'ev's Comment \cite{Comment} that they
have proposed ,,a new theoretical treatment of the temperature dependence
of the diffusion of isotopic impurities in solid $^4$He" as well as that
my theory of phonon scattering ,,fails to fit the data by very
large factors``. The latter assertion is very important because it concerns the
generally accepted selfconsistent approach to the diffusion in quantum
crystals (see e.g.
\cite{P1970,PhD,Pisma,FNT,P1975,Singapore,Nauka,DSc,G1997,Allen,Mikheev,Moscow}).
It has been used as the only argument to neglect fully the phonon-impurity
scattering mechanism, and turn back to the vacancy controlled impurity
diffusion.

      KS have evaluated in \cite{KS93} the factor $A$ in the temperature
dependence of the diffusion coefficient $D = A T^{-9}$. They have found
$A$ "to be in the range $\bf A \sim 10^{-4} - 10^{-5}$"
while my theoretical prediction \cite{Pisma,FNT,P1975,PhD,Singapore,Nauka}
is ${\bf A \sim 10^{-7}} cm^2s^{-1}K^{-9} $  and the experimental
values are $\bf A \approx 10^{-7}$ (Ref. \cite{Allen}) or $\bf A\approx
2.4\times 10^{-7}$ (Ref. \cite{Moscow}). They concluded that  "the phonon
scattering
is too weak by a factor of at least 100 to explain the observed diffusion"
(\cite{KS93}, p. 16579). KS have used in
their calculation eqs. (15) and (16) given below as (1) and (2) respectively:
\be
D = \frac{1}{3} z a^2 J^2 \tau \approx 4.3\times 10^{-15} J^2 \tau,
\ee
\be
\tau =\left[ \frac{96 \pi^9}{5} (6 \pi^2)^{2/3}s^2\right]^{-1}
\left(\frac{\Theta_D}{T}\right)^9 \omega_D^{-1} \approx 0.79 \times
10^{-17}\Theta_D^{8} T^{-9}
\ee
where $\Theta_D$ is the Debye temperature, $\omega_D= k \Theta_D/\hbar$,
$ a = 3.27\times 10^{-8}$ cm is the interatomic distance, $J$
is the tunneling frequency and $s = 1/3$. Hence,
\be
 A = 3.4\times 10^{-32} J^2\Theta_D^8
\ee
Although these expressions do not exactly coincide with those used in my works
they should give a correct order of magnitude for the diffusion coefficient
and, in particular, the order of $A$.
KS have obtained for $\Theta_D = 30 $K and $J = 2.5$ MHz the value ${\bf A =
6.0\times 10^{-5}} \, cm^2s^{-1}T^9$ while the right value (obtained by
substituting the same numbers into (3)) is  $\bf A = 1.4\times 10^{-7}$.
 The issue is not that the above expressions are the most fundamental or give
the best value of $\bf A$, but rather that they yield $A$ values consistent
with experiment. The value obtained by KS and leading to the
rejection of the phonon scattering mechanism (cf \cite{KS93}, p. 16579) may
come, therefore, from  an arithmetical error.

      In fact, $\Theta_D = 30 $ K is the upper limit and corresponds to a molar
volume $V_m = 19.8 $\, cm$^3$ while the experimental values cited above are for
$V_m = 21$ \, cm$^3$ with $\Theta_D = 26 $ K  \cite{Debye}. This yields
$\bf A = 0.44\times10^{-7}$. An ambiguity appears with the notation $J = 2.5$
MHz (p. 16579) and $J_{34}/2\pi = 2.3\times 10^5 \, s^{-1}= 0.23$ MHz (p.
16580) used for one and the same quantity.
I suppose that $J$ is given in rad/s. Otherwise, it should have been written in
the form $J/2\pi = 2.5$ MHz (as for $J_{34}$) and would be more than 10 times
larger than $J_{34}$ leading to $J = 2\pi 2.5\times 10^6\hbar/k_B = 1.2\times
10^{-4}$ K and to an energy band width $\Delta = 2zJ \approx 3\times 10^{-3}$ K
in a drastic disagreement with all known experiments. On the other
hand the value $J_{34} \approx 1.10\times 10^{-5}$ K is typical of the
exchange integral of $^3$He atoms in solid $^4$He and its substitution into (3)
gives the correct order of magnitude for $A$. Finally, even if the unrealistic
value
$J = 2\pi 2.5 \approx 15.7$ MHz were used in (3) the result for $A$ differs
from that of KS by more than an order of magnitude.

        It is clearly seen, therefore, that expressions (1) and (2) (not
necessarily the best ones) are in agreement with the experimental data. This
can be easily
verified by substituting $J$ and $\Theta_D$. Therefore, the argument that
,,the diffusion constant calculated for this theory
fails to fit the experimental data by a factor of 100`` \cite{Reply} fails.
There are no experimental data to require any drastic change of the eqs. (1)
and (2).

        It is not accurate to cite my paper \cite{Pisma} in \cite{KS93}
(Ref.18) as
if it were in support to the values of $A$  ,,in the range $10^{-4}-10^{-5}$``.
The corresponding value is $A = 2\times 10^{-7}$  as follows from eqs.(1) and
(10) in \cite{Pisma}. It was the first correct evaluation of the impuriton band
width. It is worth noting in addition that the idea
of vacancy controlled mechanism  could not be called {\it new} because it was
considered about 20 years ago (cf e.g. \cite{Locke}) and was found not
satisfactory (cf e.g. \cite{Allen}) to explain the temperature dependence of
the diffusion coefficient.

      As a consequence, the good fit of the vacancy mechanism
reported by KS after neglecting the phonon-impuriton scattering gets at least
doubtful. There are also other circumstances which call the
impurity diffusion description by KS \cite{KS93} in question, some of them
being listed in \cite{Comment} (see also \cite{G1997}). They will be
considered elsewhere.

        I do not concern the problem of impurity diffusion in solid hydrogen as
not relevant to the concrete discussion. If the theory of KS does not work
well for helium, it obviously cannot be a ,,universal theory"
for both quantum solids as argued in \cite{Reply}.

       In addition, the work of KS \cite{KS93} suffers from a number of
inaccuracies. If the dispersion law is defined by their eq.(3), then the
energy band width in a simple lattice is $\Delta = 2z J$, not $zJ$, the circles
and the squares in Fig. 1 have to change places, the works of Landesman (their
[13]) are published in Ann. de Phys. (French), not in Ann. Phys. (N.Y.), the
number $1/3$ in
eq.(5) for the dispersion law in a bcc lattice has to be replaced by $1/2$, the
term $\cos\left(\frac{\sqrt2}{3}k_z a_0\right)$ in (6) has to be replaced by
$\cos\left(\sqrt {\frac{2}{3}}k_z a_0\right)$. The correct formulae can be
found e.g. in my works\cite{PhD,FNT,DSc,Nauka}. The problem is not only in
misprints -- KS present as their own results known for 25 years. The
velocity squared $<v^2> =18 a_0^2 J^2$ (eq.(7) in \cite{KS93}) was calculated
first by Sacco and Widom\cite{Sacco} in 1976, not in their work \cite{KS92} of
1992 as cited (it is another question whether it is the right quantity for the
problem under consideration). The way of calculation of vacancy-impurity
cross-section $\sigma_0= 1.40 a^2$ is a secret. Following the references one
sees that one of the authors has made a private communication to herself
(Ref.13 in the work of Kisvarsanyi, Runge and Sullivan \cite{KS1991} reads
"K.Runge, private communication", and Ref. 14. is an unpublished M.S. Thesis
of Kisvarsanyi).


\begin{references}
\bibitem{Allen}
A. R. Allen, M. G. Richards and J. R. Schratter,
J. Low Temp. Phys. {\bf 47} 289 (1982)
\bibitem{Locke}
D. P. Locke, J.Low Temp.Phys. {\bf 32}, 159 (1978)
\bibitem{Mikheev}
V. A. Mikheev, B. N. Essel'son, V. N. Grigor'ev and N. P. Mikhin,
Fiz.Nizkih Temp. {\bf 3} 385 (1977)
\bibitem{Moscow}
B. N. Essel'son et al, in Proc. 20-th Soviet Conf. on Low.Temp.Phys. NT-20, v.2
(1979) p.236; J.de Phys.Colloque C6, suppl.8, {\bf 39} 119 (1978)
\bibitem{Pisma}
D. I. Pushkarov,
Pisma Zh.Eksp.Teor.Fiz. {\bf 19}, 751 (1974)
 [JETP Lett. {\bf 19}(12) 386-387 (1974)]
\bibitem{FNT}
D. I. Pushkarov, Fiz.Nizkih Temp. {\bf 1} 581 (1975); ibid {\bf 1} 585 (1975)
\bibitem{P1975}
D. I. Pushkarov,
Zh.Eksp.Teor.Fiz. {\bf 68} 1471 (1975);
[Sov.Phys.JETP {\bf 41 }(4)  735-737 (1976)]
\bibitem{KS93}
E. G. Kisvarsanyi and N. S. Sullivan, Phys.Rev.B {\bf 48} 16557 (1993)
\bibitem{Reply}
E. G. Kisvarsanyi and N. S. Sullivan, Phys.Rev.B {\bf 55} 3989 (1997)
\bibitem{Comment}
V. N. Grigor'ev, Phys.Rev.B {\bf 55} 3987 (1997)
\bibitem{P1970}
D. I. Pushkarov,
Zh.Eksp.Teor.Fiz. {\bf 59}, 1755 (1970)
 [Sov.Phys.JETP {\bf 32}(5) 954 (1971)]
\bibitem{PhD}
 D. I. Pushkarov - {\it "Quantum Theory of Defects in Crystals  at  Low
Temperatures"},
PhD Thesis,  Moscow  University,  Moscow,  1972  (in Russian).
\bibitem{Singapore}
 D. I. Pushkarov, {\it Quasiparticle Theory of  Defects  in
Solids},
World Scientific, Singapore-New Jersey-London-Hong Kong, 1991.
\bibitem{Nauka}
 D. I. Pushkarov,  {\it Defektony v Kristallakh \ldots}
(Defectons  in  Crystals.
Quasiparticle Approach to Defects in Solids), Moscow, Nauka 1993. (in
Russian.); [first edition, JINR P17-87-177 (1987), Dubna, Russia].
\bibitem{DSc}
 D. I. Pushkarov, {\it Quasiparticle  Approach  in  Quantum  Theory
of Solids}, DSc Thesis, Dubna, USSR, 1986. (in Russian.); Preprint JINR
17-86-223 (1986), Dubna, Russia
\bibitem{G1997}
V. N. Grigor'ev, Fiz.Nizkih Temp. {\bf 23} 5 (1997)
\bibitem{Debye}
D. O. Edwards and R. C. Pandorf, Phys.Rev. {\bf 140}, A816 (1965)
\bibitem{Sacco}
J. E. Sacco, A. Widom, J.Low Temp.Phys. {\bf 24} 241 (1976)
\bibitem{KS92}
M.Rall, D.Zhou, E.K.Kisvarsanyi and N.S.Sullivan, Phys.Rev.B45, 2800 (1992)
\bibitem{KS1991}
E. G. Kisvarsanyi, K. Runge and N. S. Sullivan, Phys.Lett.A {\bf 155} 337
(1991)
\end{references}
\end{document}